\documentclass{iau}
\usepackage{graphicx}
\usepackage{amsmath}
\usepackage{array}
\usepackage{booktabs}
\usepackage[labelfont=bf]{caption}
\title[Kinetic Monte Carlo simulations of the grain-surface back-diffusion effect] 
{Kinetic Monte Carlo simulations of the grain-surface back-diffusion effect}

\author[Willis \& Garrod]   
{Eric R. Willis$^1$ \and Robin T. Garrod$^{1,2}$}

\affiliation{$^1$Department of Chemistry, University of Virginia \\ Charlottesville, VA
22904-4319 \\ email: {\tt ew2zb@virginia.edu} \\[\affilskip]
$^2$Department of Astronomy, University of Virginia \\ Charlottesville, VA
22904-4325}

\pubyear{2017}
\volume{332}  
\jname{Astrochemistry VII: Through the Cosmos from Galaxies to Planets}
\editors{M. Cunningham, T.Millar \& Y. Aikawa, eds.}
\begin{document}

\maketitle

\begin{abstract}
Back-diffusion is the phenomenon by which random walkers revisit binding sites on a lattice. This phenomenon must occur on interstellar dust particles, slowing down dust-grain reactions, but it is not accounted for by standard rate-equation models. Microscopic kinetic Monte Carlo models have been used to investigate the effect of back-diffusion on reaction rates on interstellar dust grains. Grain morphology, size, and grain-surface coverage were varied and the effects of these variations on the magnitude of the back-diffusion effect were studied for the simple H+H reaction system. This back-diffusion effect is seen to reduce reaction rates by a maximum factor of $\sim$5 for the canonical grain of 10$^6$ binding sites.The resulting data were fit to logarithmic functions that can be used to reproduce the effects of back-diffusion in rate-equation models.
\end{abstract}

\firstsection 
\section{Introduction}
Astrochemical rate-equation models are widely-used to investigate the chemistry in interstellar environments, from molecular clouds to protoplanetary disks. Despite their accuracy in reproducing gas-phase chemistry, these models have some documented deficiencies with regards to grain-surface chemistry (e.g., Garrod 2008). While they can not be used to simulate large systems like rate-equation models, microscopic Monte Carlo models can supply accurate and detailed information about grain-surface chemistry (Garrod 2013, Pederson \textit{et al.} 2014). This information can then be used to improve rate-equation models by, for example, incorporating physical effects (such as back-diffusion) that are otherwise difficult to analytically solve. 

\section{Methods}
In our study, we have utilized two Monte Carlo kinetics models. The simpler of the two models consists of a 2-dimensional square lattice with periodic boundary conditions. Each binding site on this lattice has rectangular geometry. For simplicity, only one particle is allowed to diffuse for each model run. This is analogous to having one mobile H atom and numerous relatively immobile O atoms. The size of the surface was varied to test surface size effects, as was the number of stationary reactive particles.

The other model used in this investigation was MIMICK (Model for Interstellar Monte Carlo Ice Chemical Kinetics, Garrod 2013). MIMICK is a fully 3-dimensional code, and as such is capable of incorporating more complex grain geometries. For the purposes of this study, two geometries were used: a cubic grain and a bucky-ball. These different geometries were chosen in order to study the effect of large-scale and small-scale geometric changes on the back-diffusion effect. To this end, the cubic grain has rectangular binding site geometry (as the simple 2-dimensional model does), whereas the bucky-ball grain has hexagonal binding site geometry. In all MIMICK models, a single diffusion barrier is used for all binding sites (510.6 K, taken from Lohmar \& Krug 2009), and a simplified chemical network including only atomic and molecular hydrogen is included. When two H atoms meet on the surface, they are assumed to immediately react to form H$_2$, which then desorbs from the surface. 

\section{Results}
The primary results of this study are the data obtained with MIMICK for the cubic and bucky-ball 3-dimensional grains. These data are shown in Figures 1 and 2, and plot the back-diffusion factor ($\phi$) varying with grain surface coverage. The back-diffusion factor is simply the factor by which grain-surface reactions are slowed compared to a standard rate-equation treatment.

Grain-surface coverage has the largest effect on the back-diffusion factor. The behavior can be divided into 3 basic regimes. At high coverage, $\phi$ approaches 1, indicating that there is a high probability of reaction within a few hops. At intermediate surface coverage, $\phi$ can be described by a logarithmic function dependent upon surface coverage. At very low surface coverage, the back-diffusion factor reaches a maximum at 2 particles on the grain, with a plateau value that can be fit using a logarithmic function dependent on surface size.

\begin{figure}[h!]
   \centering
   \includegraphics[width=.9\linewidth]{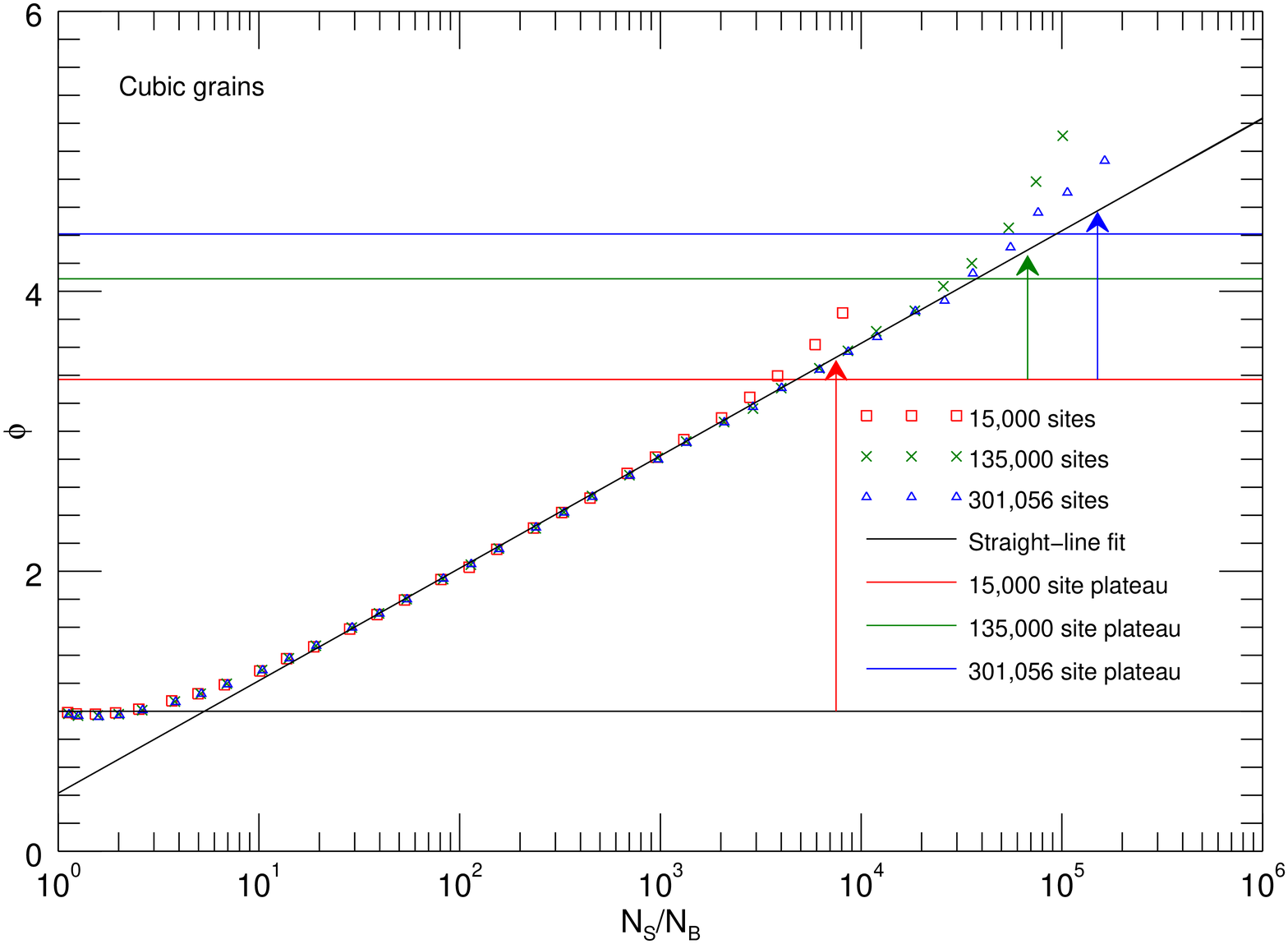}
   \caption{Results for cubic grains.. The arrows are color-coded to correspond to the data points for the same grain size, and indicate the points on the line corresponding to an average of two particles on each grain.}
\end{figure}

\begin{figure}[h!]
   \centering
   \includegraphics[width=.9\linewidth]{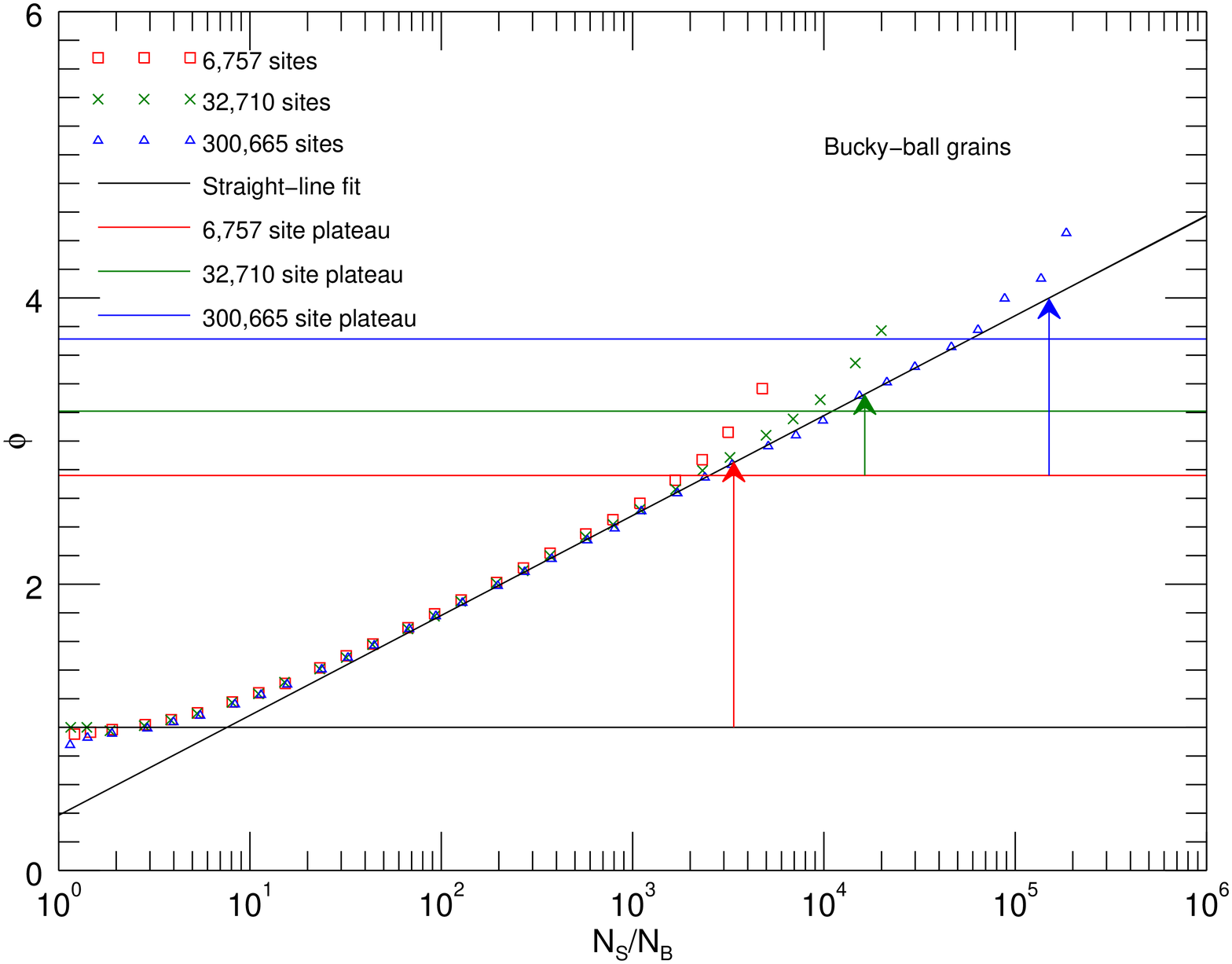}
   \caption{Results for bucky-ball grains, as per Fig. 1.}
\end{figure}

Smaller effects are observed for grain morphology and size. For each grain morphology, the different sizes are observed to follow the same logarithmic trend until a certain point of low coverage is achieved, where the grains deviate from the logarithmic fit. This deviation is due to ``dead time'' spent in the N=0 and N=1 population states for these low-coverage models. To account for this, simulations were run limiting the number of particles on the grain to 2. The back-diffusion factor calculated from these simulations is assumed to be a maximum, and is displayed as the color-coded horizontal lines in the figures. 

The cubic and bucky-ball grain morphologies exhibit slight differences in their back-diffusion factors with coverage, corresponding to the aforeenetioned differences in local binding site geometry. It is observed that the bucky-ball grain with hexagonal binding site geometry produces lower values of $\phi$ for identical coverages. This effect is in agreement with previous simulations (Chang, Cuppen \& Herbst 2005).

\begin{table}[h!]
   \centering
   \setlength{\aboverulesep}{0pt}
   \setlength{\belowrulesep}{0pt}
   \caption{Fits for each grain morphology. Straight-line fit corresponds to the main logarithmic portion of the data, while plateau fit corresponds to the maximum value for each grain size.}
   \begin{tabular}{|c|c|c|}
      \toprule
      Morphology & Straight-line fit & Plateau fit \\
      \midrule
      Flat (single mobile particle) & 0.315 $\ln{\frac{N_{S}}{N_{B}}}$ + 0.950 & 0.315 $\ln{N_{S}}$ + 0.200 \\
      \midrule
      Cubic  & 0.349 $\ln{\frac{N_{S}}{N_{B}}}$ + 0.415 & 0.342 $\ln{N_{S}}$ + 0.072 \\
      \midrule
      Bucky-ball  & 0.303 $\ln{\frac{N_{S}}{N_{B}}}$ + 0.386 & 0.250 $\ln{N_{S}}$ + 0.580 \\
      \bottomrule
   \end{tabular}
\end{table}

The logarithmic fits for $\phi$ were tested in a simple rate-equation model. These fits were seen to reproduce the back-diffusion behavior observed in the Monte Carlo models quite accurately, with $>$95\% agreement at almost all coverages. The fits are simple to incorporate into rate-equation models, and should be inserted into the computation of the rate coefficients on the grain surface as follows, where $\phi$ is the back-diffusion factor:

\begin{equation} \tag{1}
    k_{reac}=\frac{k_{hop,i}+k_{hop,j}}{\phi N_s}
\end{equation}
The fits for all surface morphologies are shown in Table 1. If the calculated back-diffusion factor is less than 1 or greater than the plateau for a particular grain size, it should be corrected to these values. 
\\

\noindent The authors would like to acknowledge Eric Herbst, Aspen R. Clements, and Ilsa R. Cooke for useful discussions. This work was supported by the NASA Laboratory Astrophysics Program (Grant NNX15AG07G).

\end{document}